\documentclass[11pt]{article}

\let\counterwithin\relax

\usepackage{amsmath, amsfonts, amssymb, amsthm, bm, graphicx, mathtools, enumerate,multirow}
\usepackage[sort&compress]{natbib}
\usepackage[CJKbookmarks=true,
bookmarksnumbered=true,
bookmarksopen=true,
colorlinks=true,
citecolor=blue,
linkcolor=blue,
anchorcolor=blue,
urlcolor=blue,
breaklinks=true]{hyperref}
\usepackage[usenames]{color}
\usepackage{breakcites}
\usepackage[letterpaper, left=1.2truein, right=1.2truein, top = 1.2truein,
bottom = 1.2truein]{geometry}
\usepackage[ruled, lined, commentsnumbered]{algorithm2e}
\usepackage{prettyref,soul}

\usepackage{apptools} %Ya
\usepackage{chngcntr} % define \counterwithin
\AtAppendix{\counterwithin{lemma}{section}} %Ya
\usepackage{tikz}
\usepackage[font=small,labelfont=bf]{caption}
\usepackage{enumitem}

\usepackage[caption=false,font=normalsize,labelfont=sf,textfont=sf]{subfig}

\usepackage{booktabs} % 提供 \toprule \midrule \bottomrule 等命令
\usepackage{multirow} % 提供 \multirow 跨行功能
\usepackage{makecell}
\usepackage{siunitx}
\usepackage{adjustbox}
\usepackage{longtable}

\theoremstyle{definition}

\newrefformat{eq}{(\ref{#1})}
\newrefformat{chap}{Chapter~\ref{#1}}
\newrefformat{sec}{Section~\ref{#1}}
\newrefformat{algo}{Algorithm~\ref{#1}}
\newrefformat{fig}{Fig.~\ref{#1}}
\newrefformat{tab}{Table~\ref{#1}}
\newrefformat{rmk}{Remark~\ref{#1}}
\newrefformat{clm}{Claim~\ref{#1}}
\newrefformat{def}{Definition~\ref{#1}}
\newrefformat{cor}{Corollary~\ref{#1}}
\newrefformat{lmm}{Lemma~\ref{#1}}
\newrefformat{lemma}{Lemma~\ref{#1}}
\newrefformat{prop}{Proposition~\ref{#1}}
\newrefformat{app}{Appendix~\ref{#1}}
\newrefformat{ex}{Example~\ref{#1}}
\newrefformat{cond}{Condition~\ref{#1}}

\usepackage{float}
\usepackage{listings}
\usepackage{xcolor}

\usepackage{threeparttable}

\lstdefinestyle{mystyle}{
	backgroundcolor=\color{backcolour},   
	commentstyle=\color{codegreen},
	keywordstyle=\color{magenta},
	numberstyle=\tiny\color{codegray},
	stringstyle=\color{codepurple},
	basicstyle=\ttfamily\footnotesize,
	breakatwhitespace=false,         
	breaklines=true,                 
	captionpos=b,                    
	keepspaces=true,                 
	numbers=left,                    
	numbersep=5pt,                  
	showspaces=false,                
	showstringspaces=false,
	showtabs=false,                  
	tabsize=2
}
\usepackage{makecell}

\usepackage[T1]{fontenc}
\usepackage[utf8]{inputenc}
\usepackage{authblk}
\title{
Detecting Structural Heart Disease from Electrocardiograms via a Generalized Additive Model of Interpretable Foundation-Model Predictors
}

\author[1,\#,*]{Ya Zhou}
\author[1,\#]{Zhaohong  Sun}
\author[2,\#]{Tianxiang Hao}
\author[3,*]{Xiangjie Li}

\affil[1]{Department of Information Center, Fuwai Hospital, Chinese Academy of Medical Sciences and Peking Union Medical College, Beijing, 100037, China}
\affil[2]{The Tsinghua Shenzhen International Graduate School, Tsinghua University, Shenzhen, 518055, China}
\affil[3]{National Clinical Research Center for Cardiovascular Diseases, Fuwai Hospital, Chinese Academy of Medical Sciences and Peking Union Medical College, National Center for Cardiovascular Diseases, Beijing, 100037, China}

\date{}

\begin{document}
\maketitle
\def\thefootnote{\#}\footnotetext{These authors contributed equally.}
\def\thefootnote{\arabic{footnote}}

\def\thefootnote{*}\footnotetext{Corresponding authors:Ya Zhou (zhouya@fuwai.com), Xiangjie Li (ele717@163.com) }\def\thefootnote{\arabic{footnote}}

\begin{abstract}
	Structural heart disease (SHD) is a prevalent condition with many undiagnosed cases, and early detection is often limited by the high cost and accessibility constraints of echocardiography (ECHO). Recent studies show that artificial intelligence (AI)-based analysis of electrocardiograms (ECGs) can detect SHD, offering a scalable alternative. However, existing methods are fully black-box models, limiting interpretability and clinical adoption. To address these challenges, we propose an interpretable and effective framework that integrates clinically meaningful ECG foundation-model predictors within a generalized additive model, enabling transparent risk attribution while maintaining strong predictive performance. Using the EchoNext benchmark of over 80,000 ECG--ECHO pairs, the method demonstrates relative improvements of +0.98\% in AUROC, +1.01\% in AUPRC, and +1.41\% in F1 score over the latest state-of-the-art deep‐learning baseline, while achieving slightly better performance even with only 30\% of the training data. Subgroup analyses confirm robust performance across heterogeneous populations, and the estimated entry-wise functions provide interpretable insights into the relationships between risks of traditional ECG diagnoses and SHD. This work illustrates a complementary paradigm between classical statistical modeling and modern AI, offering a pathway to interpretable, high-performing, and clinically actionable ECG-based SHD screening.
\end{abstract}

\section{Introduction}
\label{sec:intro}
Structural heart disease (SHD) is a global health challenge for which  early detection is critical to prevent adverse outcomes \citep{poterucha2025detecting}. SHD includes abnormalities of cardiac valves, myocardium, and chambers, such as valvular heart disease and heart failure, affecting an estimated tens of millions of individuals worldwide \citep{ulloa2022rechommend, wang2023summary, tsao2023heart, mensah2023global}. Despite its prevalence, a substantial proportion of cases remain undiagnosed. For example, estimates from large community-based screening studies suggest that the prevalence of clinically meaningful valvular disease more than doubles when systematic echocardiography (ECHO) is performed, indicating a considerable burden of missed disease in routine care \citep{d2016large}. Although ECHO remains the diagnostic gold standard, its high cost and reliance on specialized expertise limit its broad accessibility \citep{kwon2020deep, diao2025speed}. Furthermore, symptoms of many SHD subtypes are nonspecific or arise only in advanced stages, narrowing the window for effective intervention \citep{poterucha2025detecting}. These challenges highlight the need for scalable, accurate, and accessible tools to identify individuals with SHD.

%Furthermore, symptoms of many SHD subtypes arise late or overlap with other conditions, opportunities for early intervention are often missed \citep{poterucha2025detecting}.

%Structural heart disease (SHD) represents a significant global health burden, and early detection is critical for improving outcomes \citep{poterucha2025detecting}. SHD includes abnormalities of cardiac valves, myocardium, and chambers, such as valvular heart disease and heart failure, affecting an estimated tens of millions of individuals worldwide, yet a substantial proportion remains undiagnosed \citep{ulloa2022rechommend, wang2023summary, tsao2023heart, mensah2023global}. For instance, a large-scale community study has shown that the prevalence of clinically significant valvular disease more than doubles when systematic echocardiography (ECHO) is performed, indicating that many cases may be missed in routine care \citep{d2016large}. Although ECHO provides a definitive diagnosis, its high cost and reliance on specialized expertise limit its broad accessibility \citep{kwon2020deep, diao2025speed}. As symptoms of at least two forms of SHD arise late or overlap with other conditions, opportunities for early intervention are often missed \citep{poterucha2025detecting}. These challenges highlight the need for scalable, accurate, and accessible tools to identify individuals with SHD. 

Electrocardiography (ECG), a widely available and low-cost test that records cardiac electrical activity, offers a scalable option for population-level screening. However, conventional clinical interpretation is generally insufficient for detecting SHD, as relevant signal patterns in ECG signals may be subtle and imperceptible to the human eye \citep{siontis2021artificial}. Recent advances in deep learning have shown that models trained on large-scale ECG datasets paired with ECHO labels can identify these subtle patterns with high predictive performance \citep{attia2019screening, cohen2021electrocardiogram}. This line of work demonstrates the feasibility of ECG-based detection of SHD and has led to growing research interest in leveraging ECG signals for earlier and scalable identification \citep{poterucha2025detecting, diao2025speed}.

However, most existing approaches are fully black-box models \citep{somani2021deep, siontis2021artificial, attia2019screening, cohen2021electrocardiogram, poterucha2025detecting, diao2025speed}, which limits both interpretability and clinical adoption. The dominant paradigm constructs end-to-end deep learning models that map ECG signals directly to ECHO-derived labels. Unlike conventional ECG tasks, such as detection of atrial fibrillation, sinus tachycardia, or sinus arrhythmia, where ECG serves as the gold standard with visible signal changes clearly described in clinical guidelines, the relevant patterns associated with SHD may be subtle and not conform to traditional ECG knowledge \citep{siontis2021artificial}. Consequently, despite strong predictive performance, these models remain difficult to interpret and raise substantial concerns regarding transparency, validation, and real-world deployment.

To address these limitations, we propose a simple yet effective framework that integrates interpretable ECG foundation-model predictors within a generalized additive model. Rather than training an end-to-end black-box classifier, we construct a model in which each input corresponds to a clinically meaningful predictor extracted from a post-trained ECG foundation model. These predictors, representing calibrated risks of traditional ECG diagnoses, are widely used in clinical practice and familiar to practicing clinicians.  Importantly, each predictor also reflects a specific rhythm or morphological pattern embedded in the underlying ECG signal, providing a physiologically grounded representation of cardiac electrical activity. Motivated by the fact that SHD-related ECG patterns can be highly complex, we employ nonparametric functional estimation to flexibly characterize their contributions while preserving transparency. This formulation enables users to examine how individual ECG-derived patterns contribute to SHD risk and to evaluate whether the inferred relationships align with established clinical knowledge. Together, this approach maintains interpretability and physiological relevance while achieving strong predictive performance.

We assess the proposed method using the EchoNext benchmark dataset \citep{elias2025echonext}, which contains more than 80,000 ECG–ECHO pairs. The predictor extractor is constructed by applying the post-training strategy from \citep{zhou2025bridging} to the ECG foundation model ST-MEM \citep{na2024guiding}, using the PTB-XL dataset \citep{wagner2020ptb} with 71 traditional ECG diagnostic labels. For the additive modeling component, we employ B-spline bases with commonly recommended spline order and basis dimension following standard practice \citep{huang2010variable, fan2014nonparametric}. 
Compared with the state-of-the-art Columbia mini model \citep{poterucha2025detecting}, the proposed approach achieves relative improvements of +0.98\% in AUROC, +1.01\% in AUPRC, and +1.41\% in F1 score. Notably, when trained using only 30\% of the available data, the proposed framework still slightly outperforms the Columbia mini model, indicating improved data efficiency. Subgroup analyses indicate that model performance remains stable across heterogeneous demographic and clinical subpopulations. Moreover, examination of the estimated entry-wise functions (i.e., component-wise nonlinear effects) reveals associations linking traditional ECG diagnostic risks to SHD, providing interpretable structure and potential clinical insight into the prediction mechanism.

In summary, the main contributions of this work are as follows:

\begin{itemize}
\item [1. ]	We propose a new interpretable modeling framework that integrates ECG foundation-model  predictors with a generalized additive structure, offering a transparent alternative to existing end-to-end black-box SHD detection methods.
\item [2. ] The proposed method achieves strong predictive performance and improved data efficiency on a benchmark of more than 80,000 ECGs, outperforming the current state-of-the-art model while maintaining interpretability and demonstrating robustness across demographic and clinical subgroups.
\item [3. ] This work establishes a complementary paradigm between statistical modeling and foundation-model AI, demonstrating that interpretability and high predictive performance can be achieved simultaneously.
\end{itemize}

The remainder of this article is organized as follows. In Section \ref{sec:model}, we introduce the proposed framework. Section \ref{sec:method} specifies the estimation methods to obtain the predictor extractor and additive components in the framework. We conduct experiments on a large-scale real-world dataset in Section \ref{sec:exp}. Additional discussions are provided in Section \ref{sec:dis}, and Section \ref{sec:con} presents a brief conclusion.

\section{Model}
\label{sec:model}

We consider fixed-length multi-lead electrocardiograms (ECGs), such as standard 12-lead recordings with a typical duration of 10 seconds. Each ECG is represented as a multivariate time series 
\[
\mathbf{X} \in \mathbb{R}^{L \times Q},
\]
where \(L\) denotes the number of leads and \(Q\) denotes the number of sampled time points. In addition, demographic or clinical covariates (e.g., age and sex) are represented as a vector \(\mathbf{z} \in \mathbb{R}^{p}\). The binary response \(y \in \{0,1\}\) indicates the presence of structural heart disease (SHD).
%, either as a composite condition or a specific subtype.

We model the conditional mean response through a generalized additive formulation:
\begin{equation}
\label{eq:model}
g\{\mathbb{E}(y \mid \mathbf{z}, \mathbf{X})\}
= \boldsymbol{\gamma}^{\top}\mathbf{z} + m(\mathbf{X}),
\end{equation}
where \(g(\cdot)\) is a fixed link function (e.g., logistic link), \(\boldsymbol{\gamma} \in \mathbb{R}^{p}\) are unknown regression parameters, and \(m(\cdot)\) is an unknown function capturing the effect of ECG morphology.

In practice, the response $y$ is derived from echocardiography (ECHO) rather than manual ECG interpretation. Since SHD-related electrical abnormalities might be subtle and unrecognizable to human eyes \citep{siontis2021artificial}, the end-to-end mapping from $\mathbf{X}$ to $y$ tends to be highly nonlinear and difficult to interpret. Thus, the primary methodological challenge is to estimate $m(\cdot)$ in a manner that is both accurate and clinically interpretable.

To motivate the proposed approach, we first review existing modeling strategies and their limitations in Sections  \ref{subsec:statistical_modeling} and \ref{subsec:deep_learning}. Section~\ref{subsec:hybrid_modeling} then introduces a simple yet effective modeling framework that provides both  predictive accuracy and clinical interpretability.

\subsection{Classical statistical modeling: interpretable but limited expressive capacity}
\label{subsec:statistical_modeling}

%A natural starting point is to assume a linear form for $m(\cdot)$. However, such a specification is overly restrictive for ECG-based SHD prediction and fails to capture nonlinear temporal or cross-lead physiological dependencies.
%A common extension is to adopt an additive representation of the form \citep{stone1985additive, hastie1990generalized}:
%where $\nu \in \mathbb R$ is an intercept term, $X_{l,q}$ is the $(l,q)$-th entry of  $\mathbf X$ and $m_{l,q}$'s are unknown univariate functions. Although this formulation allows nonlinear effects, it introduces $LQ$ separate unknown components.  For a standard 12-lead ECG sampled at 250 Hz over 10 seconds, $LQ=30,000$, making estimation difficult. Furthermore, the additive structure of univariate functions prevents the model from capturing nonlinear physiologic dependencies across both leads and time points, which are necessary for interpreting spatially and temporally coordinated electrophysiologic patterns \citep{na2024guiding}. 

Classical statistical modeling approaches provide interpretability, but they often lack sufficient expressive capacity when applied to ECG signals. This limitation typically results in suboptimal predictive accuracy.

A natural starting point is to assume a linear form for $m(\cdot)$, which is simple to estimate and provides transparent risk attribution. However, linear models are overly restrictive for ECG-based SHD prediction. To relax these restrictions, an additive modeling approach \citep{stone1985additive, hastie1990generalized} can be adopted:
\begin{equation}
\label{eqn:rough_additive_model}
m(\mathbf{X}) = \nu + \sum_{l,q} m_{l,q}(X_{l,q}),
\end{equation}
where $\nu \in \mathbb R$ is an intercept term, $X_{l,q}$ denotes the $(l,q)$-th entry of $\mathbf X$, and $m_{l,q}$ are unknown univariate functions. This formulation allows each predictor to have a flexible, potentially nonlinear effect on the response. However, for a standard 12-lead ECG sampled at 250 Hz over 10 seconds, $LQ=30,000$, resulting in a prohibitively large number of separate unknown functions to estimate. Furthermore, the additive structure of univariate functions inherently precludes modeling nonlinear physiologic dependencies across both leads and time points, which are necessary for interpreting spatially and temporally coordinated electrophysiologic patterns \citep{na2024guiding}.

In principle, higher-order models that incorporate multivariate interactions could address these limitations. Yet specifying such interactions requires substantial domain knowledge and quickly leads to combinatorial model growth, making these approaches impractical at scale, particularly for ECG data analysis. 

These limitations motivate the use of deep learning methods, which can flexibly model nonlinear temporal structure and cross-lead dependencies that traditional approaches fail to capture. However, despite their strong predictive performance, such models introduce new challenges related to transparency and interpretability, as discussed below.

\subsection{Pure deep learning: accurate but difficult to interpret}
\label{subsec:deep_learning}

Recent work has demonstrated that deep learning models trained end-to-end on raw ECG signals can achieve strong predictive performance in detecting subtle ECG abnormalities. Due to the clinical importance of SHD, end-to-end deep learning has been widely explored for predicting either composite SHD or specific subtypes from ECG data. For example, \cite{attia2019screening} trained a convolutional neural network (CNN) to detect ventricular dysfunction using paired 12-lead ECGs and ECHO-derived left ventricular ejection fraction measurements from 44,959 patients. \cite{cohen2021electrocardiogram} developed a CNN to identify moderate-to-severe aortic stenosis using 129,788 ECG–ECHO pairs. \cite{poterucha2025detecting} trained a CNN using more than one million ECG–ECHO pairs from a large integrated health system to detect SHD. More recently, \cite{diao2025speed} proposed a self-distilled and pre-trained Transformer model trained on 466,149 paired ECG–ECHO samples to detect tricuspid regurgitation and its subtypes.

Although these models achieve high predictive accuracy, they remain difficult to interpret. Clinicians cannot readily understand why an ECG is classified as positive or negative, which limits trust and clinical adoption. Post-hoc interpretability tools such as attribution maps or Grad-CAM \citep{hata2020classification} can highlight input regions influencing model decisions; however, the highlighted waveform segments often do not correspond to recognizable clinical features, especially when the underlying disease signatures are subtle.

These challenges motivate the development of modeling strategies that preserve the predictive strengths of deep learning while ensuring transparent decision-making. To address this gap, we propose a modeling framework that couples high-capacity deep ECG representations with a structured additive model, enabling clinically interpretable latent predictors without sacrificing accuracy, as discussed below.

%Moreover, such approaches are inherently sample-specific \citep{zhang2021survey}, meaning that ECGs with identical labels may produce vastly different explanation maps, raising concerns about reliability.

\subsection{A hybrid modeling framework with interpretable latent predictors}
\label{subsec:hybrid_modeling}

We propose a hybrid modeling framework that integrates deep learned ECG representations with additive statistical modeling. The central idea is to transform raw ECG waveforms into a structured set of interpretable latent predictors that reflect clinically meaningful morphologic and rhythm characteristics. The unknown function $m(\cdot)$ is modeled using an additive function of these predictors, formulated as:
\begin{equation}
\label{eqn:def:model}
m(\mathbf{X}) = \sum_{j=1}^{J} f_j[\sigma \{ h_j(\mathbf{X})\} ],
\end{equation}
where $J$ denotes the predefined number of predictors, $h_j: \mathbb{R}^{L\times Q} \to \mathbb{R}$ represents a latent predictor derived from the ECG, $\sigma(\cdot)$ is the sigmoid function and $f_j(\cdot)$ is an unknown smooth univariate function capturing the potentially nonlinear association between the predictor and structural heart disease. Unlike conventional ECG modeling frameworks where $h_j(\cdot)$ is manually engineered based on expert-defined rules, here each $h_j(\cdot)$ corresponds to a logit output of a deep neural network. In practice, a set of logits corresponding to traditional ECG diagnostic categories, such as atrial fibrillation, sinus bradycardia, or sinus arrhythmia, may be selected as interpretable predictors. After applying a sigmoid transformation, these logits yield values representing risks for clinically defined conventional ECG diagnoses, which are grounded in well-established waveform patterns and clinical guidelines. For such diagnostic tasks, deep learning models have demonstrated performance comparable or superior to cardiologists \citep{hannun2019cardiologist, ribeiro2020automatic, jiang2024self}.  Importantly, such models have been deployed clinically in many hospitals, making the resulting latent predictors familiar and interpretable to clinicians.

It is important to note that traditional ECG diagnostic tasks differ substantially from detecting SHD using ECG. The former reflects overt waveform abnormalities that are recognizable and clinically defined, whereas the latter depends on subtle physiologic signatures that might not be visually interpretable \citep{siontis2021artificial}. Nevertheless, the availability of logits corresponding to many clinically validated ECG diagnostic labels provides a rich set of latent variables that capture diverse electrophysiologic characteristics. By modeling their effects nonparametrically through the functions $f_j(\cdot)$, the proposed framework can flexibly represent nonlinear associations between SHD and these logits, while preserving interpretability at the level of individual predictors. We demonstrate these advantages using a large-scale real-world data in Section \ref{sec:exp}.

For the choice of the network $h_j(\cdot )$ serving as the predictor extractor, numerous methods have been proposed for ECG modeling \citep{strodthoff2020deep, cheng2023msw, shi2023sequence, zhang2023effectively, behrouz2024chimera, na2024guiding, li2025electrocardiogram, zhou2025enhancing}. In principle, any model capable of accurate and comprehensive ECG diagnostic prediction may serve this role. Recent ECG foundation models are particularly well suited due to their scalability, broad adaptability, and ease of use through widely available open-source implementations. In this work, we adopt such a model and apply a post-training strategy to instantiate the proposed framework.

\section{Methodology}
\label{sec:method}

Section \ref{sec:model} introduced the proposed hybrid modeling framework. In this section, we detail a concrete estimation strategy, which integrates ECG foundation-model predictor extraction with nonparametric statistical modeling. Implementing the framework involves specifying three components:

\begin{itemize}
\item[(1)] the selection of an appropriate ECG foundation model;
\item[(2)] the choice of interpretable predictors and the associated post-training strategy used to learn $h_j(\cdot)$, $j=1,\ldots,J$; 
\item[(3)] the estimation approach for the unknown functions $f_j(\cdot)$, $j=1,\ldots,J$.
\end{itemize}

We describe each component in the subsections below.

\subsection{ECG foundation model}
\label{sec:ST_MEM}
As discussed in Section \ref{subsec:hybrid_modeling}, any model capable of broad and accurate ECG diagnostic prediction can serve as the feature extractor $h_j(\cdot)$. In this work, we adopt an ECG foundation model that is scalable, broadly adaptable, and easy to use via publicly available open-source implementations.

Several ECG foundation models have been proposed in recent years, including ECG-FM \citep{mckeen2024ecg}, HuBERT-ECG \citep{coppola2024hubert}, ECGFounder \citep{li2025electrocardiogram}, and ST-MEM \citep{na2024guiding}, spanning both CNN-based and Transformer-based architectures. We use ST-MEM, a variant of the Transformer model trained within a masked autoencoder framework \citep{he2021masked}; a related variant has demonstrated robustness in large-scale real-world ECG settings \citep{diao2025speed, zhou2025enhancing}. Importantly, both the pretrained weights and implementation of ST-MEM are publicly available, enabling reproducibility and ease of use. Other foundation models may also be compatible with our framework and could be explored in future work.

\subsection{Interpretable predictors and post-training}
Traditional ECG diagnostic labels often exhibit hierarchical structure, in which fine-grained rhythm or morphology annotations can be grouped into broader diagnostic categories. For example, the SCP-ECG statements taxonomy \citep{wagner2020ptb} aggregates left ventricular hypertrophy, right ventricular hypertrophy, and related abnormalities into a higher-level hypertrophy category. To preserve diagnostic granularity and maximize information content, we utilize the lowest-level (i.e., most specific) diagnostic categories as candidate predictors. Since the number of available conventional ECG diagnostic labels is typically much smaller than the size of the training dataset (see Section \ref{sec:ptb_xl}), we retain all lowest-level labels rather than performing additional feature selection. After applying a sigmoid transformation to the corresponding logits, these outputs serve as interpretable latent predictors within the proposed hybrid framework.

To obtain reliable estimates for these predictors, we adopt a post-training strategy adapted from \citet{zhou2025bridging}, which has demonstrated substantial performance gains over standard fine-tuning procedures for ECG foundation models. The strategy consists of two stages. First, during the initialization stage, we apply linear probing to estimate the weights of the final classification layer based on publicly available pretrained representations. Second, during the regularization stage, we introduce stochastic depth and dropout to improve robustness and mitigate overfitting. This post-training procedure yields improved performance on conventional ECG diagnostic tasks, thereby enhancing the quality of the latent predictors used in the additive modeling stage of our framework.

\subsection{Nonparametric estimation}
\label{sec:nonparametric_est}
For each observation, the latent diagnostic predictors satisfy $\sigma\{h_j(\mathbf{X})\} \in (0,1)$, making them well suited to smooth nonparametric modeling. We approximate each unknown function $ f_j(\cdot) $ using a B-spline basis of order \( \zeta \):
\begin{equation}
f_j(x) \approx \sum_{k=1}^K \alpha_{j,k} b_{j,k}(x),
\end{equation}
where $b_{j,1}(x),\dots,b_{j,K}(x) $ denote the spline basis functions and $ \alpha_{j,k}$'s are unknown coefficients. Since predictor distributions may differ substantially across diagnostic signals, the spline basis specification is allowed to vary by $j$,$j=1,\ldots,J$. Substituting the expansion into \eqref{eqn:def:model} yields the spline-approximated form:
\begin{equation}
\label{eqn:def_m_approx}
m(\mathbf{X}) \approx \sum_{j=1}^J \sum_{k=1}^K \alpha_{j,k} b_{j,k}\big[ \sigma \{ h_j(\mathbf{X}) \} \big].
\end{equation}
Since the spline bases satisfy $ \sum_{k=1}^K b_{j,k}(x) = 1$, one degree of freedom is redundant for each function $ f_j(\cdot)$. To ensure identifiability, we remove one basis function per predictor and include a global intercept. Ignoring approximation error, the model \eqref{eq:model} can be expressed as:
\begin{equation}
\label{eqn:def_m_equal}
g\{\mathbb{E}(y \mid \mathbf{z}, \mathbf{X})\}
=   \nu +  \boldsymbol{\gamma}^{\top}\mathbf{z}  + \sum_{j=1}^J \sum_{k=1}^{K-1} \alpha_{j,k} b_{j,k}\big[ \sigma \{ h_j(\mathbf{X}) \} \big],
\end{equation}
where $g(\cdot)$ is a fixed link function,  $\nu \in \mathbb{R}$ is an intercept and $\boldsymbol{\gamma} \in \mathbb{R}^{p}$ are unknown regression parameters.

For clarity, we focus on the logistic link of \eqref{eqn:def_m_equal} in this paper. Applying an $\ell_2$ penalty for numerical stability, the estimation problem is formulated as:

\begin{equation}
\label{eqn:def:opt}
\underset{\nu,\ \bm{\gamma},\ \alpha_{j,k}}{\arg\min}\ 
\bigg[
- \sum_{i=1}^n 
\Big\{ y_i \log(\hat{y}_i) + (1-y_i)\log(1-\hat{y}_i) \Big\} 
+ \lambda_1 \Vert \bm{\gamma} \Vert_2^2 + \lambda_2 \sum_{j=1}^{J} \sum_{k=1}^{K-1} \alpha_{j,k}^2
\bigg],
\end{equation}
where 
\[
\hat{y}_i = \sigma \bigg( \nu +  \boldsymbol{\gamma}^{\top} \mathbf{z}_i  + \sum_{j=1}^J \sum_{k=1}^{K-1} \alpha_{j,k} b_{j,k}\big[ \sigma \{ h_j(\mathbf{X}_i) \} \big] \bigg),
\]
and $\lambda_1$ and $\lambda_2$ are tuning parameters. This formulation corresponds to a penalized logistic regression model with spline-expanded predictors. The optimization problem can be efficiently solved using standard numerical implementations available in commonly used software platforms such as \texttt{Python} or \texttt{R}.

The proposed framework introduces several tuning components, including the spline order $\zeta$, the number of basis functions $K$, the knot placement strategy, and the regularization parameters $\lambda_1$ and $\lambda_2$. To balance modeling flexibility and computational simplicity, we adopt a set of established yet practical choices. Specifically, following common convention, we set the spline order to $\zeta = 4$, yielding a cubic B-spline basis \citep{huang2010variable}. The number of basis functions is selected as 
$
K =  \lceil 2 n^{1/5} \rfloor,
$
where $\lceil\cdot\rfloor$ denotes rounding to the nearest integer, following recommendations in the nonparametric regression literature \citep{fan2014nonparametric}. Knots are generated in a data-driven manner using equally spaced empirical quantiles to ensure robustness under varying feature distributions. For regularization, we set $\lambda_1 = \lambda_2$ to reduce the hyperparameter search dimension. A small grid search is performed, and the final configuration is selected based on the highest Area Under the Receiver Operating Characteristic Curve (AUROC) score on a held-out validation set.

\section{Experiments}
\label{sec:exp}
%In this section, we applied the proposed method in a large scale real data to show its effectiveness. We firstly introduce two datasets, the traditional ECG dataset, used to post-train the foundation model to obtain the feature extractor, and the ECG-ECHO dataset, used to evaluate the proposed method in detection of SHD. We then report the proposed method with different settings on the ECG-ECHO dataset.

This section evaluates the proposed method using large-scale real-world ECG datasets. We first introduce the two datasets utilized in this study: a traditional ECG dataset used to post-train the foundation model to obtain a strong predictor extractor, and the ECG–ECHO dataset used to assess the proposed method for structural heart disease (SHD) detection. We then investigate model performance under different settings on the ECG–ECHO dataset.

\subsection{Datasets}
\subsubsection{Traditional ECG dataset}
\label{sec:ptb_xl}
We employ PTB-XL as the source dataset to obtain the final weights of $h_j(\cdot), j=1,\ldots, J$. PTB-XL is one of the most widely adopted benchmark resources for ECG modeling due to its dataset size, rich annotation structure, and standardized evaluation protocol \citep{wagner2020ptb, strodthoffopen}. It contains 21,837 12-lead ECG recordings, each 10 seconds in duration, collected from 18,885 unique patients. Each recording is annotated with 71 traditional ECG diagnostic categories (see Table \ref{tab:clinical_ecg}) by up to two cardiologists. The dataset includes predefined training, validation, and test splits. In our study, PTB-XL is only employed for the post-training phase of the ECG foundation model.

\begin{table}[!htbp]
\caption{
Clinical variables and traditional ECG diagnostic categories used in the proposed model. 
The  7 preprocessed clinical variables and the estimated risks associated with 71 clinically defined ECG diagnostic categories are incorporated into the generalized additive model defined in \eqref{eqn:def_m_equal}. 
Diagnostic category abbreviations follow the SCP-ECG terminology; see \citet{wagner2020ptb} for full definitions.
}
\label{tab:clinical_ecg}
\centering
\begin{tabular}{p{0.25\linewidth}p{0.6\linewidth}}
\toprule
{7 clinical variables} & {71 traditional ECG diagnostic categories} \\
\midrule
sex, ventricular rate, atrial rate, PR interval, QRS duration, QT interval, patient age at time of ECG& 
NORM, LVOLT, SR, SBRAD, ABQRS, IMI, SARRH, AFLT, AFIB, NDT, NST\_, DIG, LVH, LPFB, LNGQT, LAFB, RAO/RAE, IRBBB, RVH, IVCD, LMI, ASMI, AMI, 1AVB, ISCAL, STACH, ISC\_, PACE, ISCLA, SEHYP, ISCIL, ILMI, PVC, CRBBB, CLBBB, ANEUR, ALMI, ISCAS, TAB\_, HVOLT, LOWT, STD\_, PAC, EL, NT\_, QWAVE, INVT, LPR, VCLVH, LAO/LAE, ILBBB, ISCIN, SVTAC, INJAL, INJAS, IPMI, WPW, ISCAN, INJLA, BIGU, TRIGU, IPLMI, 3AVB, INJIL, 2AVB, PRC(S), PSVT, PMI, STE\_, INJIN, SVARR \\
\bottomrule
\end{tabular}
\end{table}

\subsubsection{ECG-ECHO Dataset}
\label{sec:echo_next}
We evaluate the proposed method using the EchoNext dataset \citep{elias2025echonext}, which contains 100,000 de-identified paired ECG–ECHO records from 36,286 unique patients at Columbia University Irving Medical Center. Each record includes a standard 12-lead ECG waveform sampled at 250 Hz with 10s duration. The demographic information and ECG-derived tabular measurements includes age, sex, ventricular and atrial rates, PR interval, QRS duration, and QT interval, as shown in Table \ref{tab:clinical_ecg}. Consistent with prior work \citep{poterucha2025detecting}, these 7 clinical variables are preprocessed and incorporated into our modeling pipeline in addition to the raw ECG signals.

Structural heart disease (SHD) labels are derived from ECHO reports and provided as a binary outcome. As mentioned in Section \ref{sec:intro}, SHD includes multiple abnormalities of cardiac valves, myocardium, and chambers, such as valvular heart disease and heart failure. Prior work has shown that model precision is influenced by the prevalence of individual SHD subtypes, and that using a composite endpoint might improves predictive precision without increasing clinical burden, since patients with high predicted risk would be triaged to ECHO regardless \citep{poterucha2025detecting}. Accordingly, we use a composite label indicating moderate or severe SHD as the prediction target.

Dataset usage follows the official inclusion and exclusion rules \citep{elias2025echonext}. The data are partitioned into training, validation, and test subsets, ensuring that all ECGs from a given patient appear in only one split to prevent data leakage. The training set may include multiple ECGs per patient to support model learning, whereas the validation and test sets retain only the most recent ECG per patient to provide an unbiased estimate of generalization performance. After applying the official filtering criteria, the final dataset consists of 82,543 ECGs. Table \ref{tab:num_echonext} summarizes the characteristics of each dataset split.

\begin{table}[!htbp]
\centering
\caption{
Patient characteristics across training, validation, and test splits in the ECG–ECHO dataset \citep{elias2025echonext, poterucha2025detecting}. Values are shown as counts and percentages. 
}
\label{tab:num_echonext}
\begin{tabular}{l|c|c|c}
\toprule
& {Training set} & {Validation set} & {Test set} \\
\midrule
{Patients (n)} & 26,218 & 4,626 & 5,442 \\
{ECGs (n)} & 72,475 & 4,626 & 5,442 \\
\midrule
\multicolumn{1}{l}{ {Age groups } } 	 \\
\midrule
\hspace{0.5em}  18--59 & 29,783 (41.1\%) & 1,787 (38.6\%) & 2,124 (39.0\%) \\
\hspace{0.5em}  60--69 & 18,745 (25.9\%) & 1,093 (23.6\%) & 1,318 (24.2\%) \\
\hspace{0.5em}  70--79 & 14,898 (20.6\%) & 975 (21.1\%) & 1,154 (21.2\%) \\
\hspace{0.5em}  80+ & 9,049 (12.5\%) & 771 (16.7\%) & 846 (15.5\%) \\
\midrule
\multicolumn{1}{l}{ {Sex}  }  \\
\midrule
\hspace{0.5em}  Female & 33,524 (46.3\%) & 2,356 (50.9\%) & 2,731 (50.2\%) \\
\hspace{0.5em}  Male & 38,951 (53.7\%) & 2,270 (49.1\%) & 2,711 (49.8\%) \\
\midrule
\multicolumn{1}{l}{{Race/ethnicity} } \\
\midrule
\hspace{0.5em}  Hispanic & 22,806 (31.5\%) & 1,351 (29.2\%) & 1,649 (30.3\%) \\
\hspace{0.5em}  White & 21,289 (29.4\%) & 1,385 (29.9\%) & 1,569 (28.8\%) \\
\hspace{0.5em}  Black & 11,559 (15.9\%) & 728 (15.7\%) & 846 (15.5\%) \\
\hspace{0.5em}  Asian & 2,602 (3.6\%) & 134 (2.9\%) & 153 (2.8\%) \\
\hspace{0.5em}  Other & 5,272 (7.3\%) & 380 (8.2\%) & 457 (8.4\%) \\
\hspace{0.5em}  Unknown & 8,947 (12.3\%) & 648 (14.0\%) & 768 (14.1\%) \\
\midrule
\multicolumn{1}{l}{{Clinical context} }	  \\
\midrule
\hspace{0.5em}  Emergency & 22,811 (31.5\%) & 1,688 (36.5\%) & 1,971 (36.2\%) \\
\hspace{0.5em}  Inpatient & 34,906 (48.2\%) & 1,903 (41.1\%) & 2,203 (40.5\%) \\
\hspace{0.5em}  Outpatient & 12,423 (17.1\%) & 858 (18.5\%) & 1,059 (19.5\%) \\
\hspace{0.5em}  Procedural & 2,335 (3.2\%) & 177 (3.8\%) & 209 (3.8\%) \\
\bottomrule
\end{tabular}
\end{table}

\subsection{Experimental settings}
\subsubsection{Post-training for the ECG foundation model}
As mentioned in Section \ref{sec:ST_MEM}, we adopt ST-MEM as the backbone for the predictor extractor $h_j(\cdot), j = 1, \ldots, J$. We adapt the post-training strategy proposed in \cite{zhou2025bridging} on the PTB-XL with 71 traditional ECG diagnoses labels. Specifically, the post-training strategy include two stages: the initialization stage and the regularization stage. In the initialization stage, we adopt the same linear probing configuration as in \citep{na2024guiding}, training for 100 epochs based on the publicly available weights. In the regularization stage, we retain the fine-tuning hyperparameters from \citep{na2024guiding}, but incorporate a stochastic depth rate of 0.1 and a dropout rate of 0.01 to enhance robustness and prevent overfitting. We use the average binary cross entropy for those 71 ECG diagnostic labels as the loss function.  The model is evaluated on the validation set after each epoch, and the checkpoint with the highest validation macro AUROC is adopted. For clarity, we represent the final predictor extractor as 
\begin{equation}
\label{eqn:def_hat_h}
\hat{\bm h} (\mathbf X) = \big ( \sigma\{ \hat{h}_1(\mathbf X) \}, \cdots,  \sigma \{\hat{h}_J(\mathbf X)\} \big ),
\end{equation}
where $J=71$ in this setting. Due to the auxiliary position, we simply report the AUROC of the post-training method on the benchmark test set, i.e., 0.945, which is significantly better than previous small deep learning models \citep{strodthoff2020deep}.

\subsubsection{The proposed generalized additive model}
Based on the predictor extractor defined in \eqref{eqn:def_hat_h} and the input ECG waveform signals, we obtain $J=71$ predictors for each sample. These predictors, together with the 7 preprocessed clinical variables described in Section \ref{sec:echo_next}, are incorporated into the optimization problem formulated in \eqref{eqn:def:opt} to estimate the final SHD detection model. As outlined in Section \ref{sec:nonparametric_est}, we employ a commonly used hyperparameter selection strategy in statistical communities. Specifically, the regularization parameters are set as $\lambda=\lambda_1=\lambda_2$, and the candidate grid is chosen as {0.001,0.01,1,10,100,1000}.
Following the evaluation protocol of \cite{poterucha2025detecting}, model performance is assessed using the area under the receiver operating characteristic curve (AUROC), the area under the precision-recall curve (AUPRC), and the F1 score. For each metric, 95\% confidence intervals are computed based on 1000 bootstrap samples. 

The proposed model is trained on the full training dataset and evaluated on the entire test set. To explore the property of the proposed method, subgroup analyses are conducted across different groups, including age groups, sex, race and ethnicity, and clinical context. Additionally, we investigate the sample efficiency of the proposed method by varying the training sample size; see Section \ref{sec:res}.

\subsection{Results}
\label{sec:res}

\subsubsection{Performance comparison}
We compare the proposed generalized additive model to five approaches:
\begin{itemize}
\item [(i)] penalized logistic regression using 7 clinical variables (Logistic regression–1),
\item [(ii)] penalized logistic regression using 7 clinical variables and the 71 foundation-model predictors defined in \eqref{eqn:def_hat_h} (Logistic regression–2), 
\item[(iii)] support vector machine using 7 clinical variables (Support vector machine-1), 
\item[(iv)] support vector machine using 7 clinical variables and the 71 foundation-model predictors defined in \eqref{eqn:def_hat_h} (Support vector machine–2), and
\item[(v)] the Columbia mini model \citep{poterucha2025detecting}, representing the current state-of-the-art.
\end{itemize}
For a fair comparison, all penalized regression models are tuned over an identical grid of $l_2$ penalty values. Both support vector machine models use a radial basis function kernel, with the kernel width parameter fixed to the default adaptive setting (\texttt{gamma="scale"}) in \texttt{scikit-learn}, and the regularization parameter $C$ tuned over the same grid as the logistic regression models. The Columbia mini model is evaluated using the publicly released trained weights \citep{poterucha2025detecting}. For brevity, we refer to the proposed approach as the additive model in the remainder of this section.

Table \ref{tab:model_performance} summarizes the comparative results. The proposed additive model achieves the strongest performance across all three evaluation metrics, yielding an AUROC of 82.8, an AUPRC of 79.7, and an F1 score of 71.8. Relative to the Columbia mini model, these improvements correspond to gains of 0.98\%, 1.01\%, and 1.41\%, respectively. Notably,  Logistic regression–2 and Support vector machine-2 exhibit substantial performance improvements over Logistic regression–1 and Support vector machine-1, respectively, highlighting the predictive utility of the foundation-model-derived predictors.

\begin{figure}[!htbp]
\centering
\includegraphics[width=14cm]{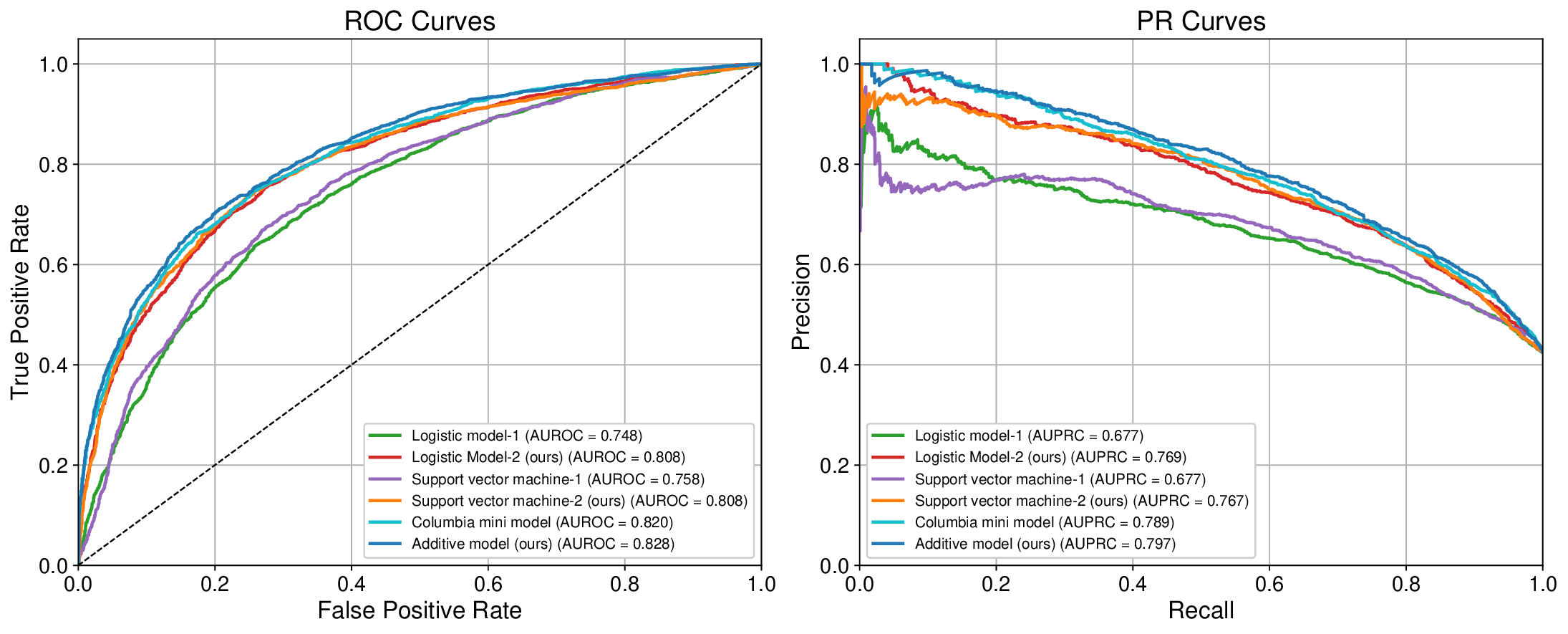}
\caption{ROC and PR curves comparison of four different models. 
The left panel shows ROC curves with AUROC values, while the right panel displays PR curves with AUPRC values. Models marked with ``(our)'' indicate models that use predictors derived from the ECG foundation model employed in this study.}
\label{fig:roc_pr}
\end{figure}

To further evaluate model performance across a range of decision thresholds, Figure \ref{fig:roc_pr} presents the receiver operating characteristic (ROC) curve and the precision–recall (PR) curve. Across both metrics, the additive model curve lies uniformly above those of all competing methods for most thresholds, confirming consistent superiority rather than performance gains limited to a particular operating point.

We additionally investigate sample efficiency by training the additive model using progressively larger fractions of the available training set. Results, also reported in Table \ref{tab:model_performance}, demonstrate monotonic improvements in AUROC, AUPRC, and F1 score as the training size increases from 10\% to 100\%. Remarkably, even with only 10\% of the training data, the additive model attains an F1 score comparable to that of the fully trained Columbia mini model. When trained on only 30\% of the data, the additive model matches or exceeds the performance of the Columbia Mini Model trained on the full dataset. These findings underscore the strong sample efficiency and robustness of the proposed approach.

\begin{table}[!htbp]
\centering
	\footnotesize
\setlength{\tabcolsep}{5pt}{
\caption{Performance metrics for the proposed additive model evaluated at varying training data sizes, alongside baseline methods. All values are reported with 95\% bootstrap confidence intervals in parentheses. Models marked with ``(our)'' indicate models that use predictors derived from the ECG foundation model employed in this study. Best performance for each metric is highlighted in bold.}
\label{tab:model_performance}
\begin{tabular}{l|c|ccc}
\toprule
Model & Training data & AUROC & AUPRC & F1 score \\
\midrule
Logistic regression-1 & 100\% & 74.8 (73.6--76.0) & 67.7 (65.6-69.8)  & 66.3 (65.0-67.7) \\
Support vector machine-1  & 100\% & 75.8 (74.5-77.0) & 67.8 (65.6-69.8)& 66.3 (65.1-67.8) \\

Logistic regression-2 (ours) & 100\% &  80.8 (79.6--81.9) & 76.9 (74.9--78.6) & 70.9 (69.7-72.4) \\
Support vector machine-2 (ours) & 100\% & 80.8 (79.7-82.0) & 76.7 (74.7-78.4) & 71.0 (69.7-72.4)
\\
Columbia mini model & 100\% & 82.0 (80.9--83.1) & 78.9 (77.2--80.4) & 70.8 (69.5--72.2) \\
\midrule
\multirow{10}{*}{Additive model (ours)}& 10\% & 81.5 (80.4--82.6) & 77.6 (75.8--79.4) & 70.6 (69.2--72.0) \\
& 20\% & 81.9 (80.8--83.0) & 78.5 (76.7--80.1) & 71.1 (69.8--72.5) \\
& 30\% & 82.1 (81.0--83.1) & 78.9 (77.2--80.5) & 71.2 (69.9--72.5) \\
& 40\% & 82.2 (81.1--83.3) &79.0 (77.4--80.6) & 71.3 (69.9--72.7) \\
& 50\% & 82.3 (81.2--83.4) & 79.2 (77.6--80.8) & 71.2 (69.9--72.7)\\
& 60\% & 82.4 (81.3--83.5) & 79.4 (77.7--81.0) & 71.0 (69.6--72.3) \\
& 70\% & 82.5 (81.4--83.5) & 79.6 (78.0--81.1) & 71.3 (70.0--72.6) \\
& 80\% & 82.7 (81.6--83.7) & \textbf{79.7} (78.0--81.2) & \textbf{71.8} (70.5--73.2) \\
& 90\% & 82.7 (81.6--83.7) & \textbf{79.7} (78.0--81.1) & 71.6 (70.3--73.1) \\
& 100\% & \textbf{82.8} (81.7--83.8) & \textbf{79.7} (78.1--81.2) & \textbf{71.8} (70.4--73.2) \\
\bottomrule
\end{tabular}
}
\end{table}

\subsubsection{Subgroup analysis}

To evaluate model robustness and fairness across patient characteristics and clinical settings, we conducted a subgroup analysis comparing the additive model with the Columbia mini model. Subgroups were defined by age, sex, race and ethnicity, and clinical context. All performance metrics are reported using the AUROC and the AUPRC, each with corresponding 95\% confidence intervals (Table \ref{tab:subgroup_performance}).

Overall, the additive model demonstrates comparable or superior performance across most subgroups. Consistent with the evaluation design used in recent work \citep{poterucha2025detecting}, age and sex were included both as model inputs and as subgroup identifiers. In 14 of 16 subgroup comparisons, the additive model achieves higher AUROC and AUPRC, while the remaining comparisons show no substantive differences.

Performance patterns were largely stable across racial and ethnic groups. The additive model achieves the highest AUROC and AUPRC in each subgroup, including Hispanic, White, Black, Asian, and other/unknown categories. Wider confidence intervals in smaller subgroups, such as the Asian group, reflect reduced sample size rather than evidence of diminished discrimination. Results were similarly consistent across clinical care contexts. The additive model maintains performance advantages in emergency, inpatient, outpatient, and procedural settings, with the largest observed margins in outpatient and procedural encounters.

Taken together, these findings suggest that the additive model generalizes well across heterogeneous patient populations and care environments, supporting its potential utility in real-world deployment scenarios.

\begin{table}[!htbp]
\centering
	\footnotesize
\setlength{\tabcolsep}{7pt}{
\caption{Performance comparison between Columbia mini model and our additive model across demographic and clinical subgroups. All values are reported with 95\% confidence intervals in parentheses. The model marked with ``(our)'' indicate the model that uses predictors derived from the ECG foundation model employed in this study. Best performance for each metric within subgroups is highlighted in bold.}
\label{tab:subgroup_performance}
\begin{tabular}{lc|cc|ccc}
\toprule
& 	\multicolumn{1}{c}{ {} }& \multicolumn{2}{c}{{Columbia mini model}} & \multicolumn{2}{c}{{Additive model (ours)}} \\
\cmidrule(lr){3-4} \cmidrule(lr){5-6}
Subgroup & $n$ & AUROC & AUPRC & AUROC & AUPRC \\
\midrule
{Age groups} \\
\midrule
\hspace{0.5em} 18--59 & 2124 & 83.0 (81.1--84.8) & 73.7 (70.7--76.8) & \textbf{83.5} (81.7--85.4) & \textbf{74.7} (71.8--77.7) \\
\hspace{0.5em} 60--69 & 1318 & 80.6 (78.2--83.0) & 75.2 (71.6--78.7) & \textbf{81.8} (79.3--84.2)& \textbf{77.5} (73.8--81.0) \\
\hspace{0.5em} 70--79 & 1154 & 77.9 (75.1--80.5) & 80.4 (77.3--83.2) & \textbf{79.0} (76.3--81.5) & \textbf{81.2} (78.1--84.1) \\
\hspace{0.5em} 80+ & 846 & 75.9 (72.7--79.0) & \textbf{85.5} (82.7--88.3) & \textbf{76.0} (72.9--79.1) & 85.1 (82.3--88.0) \\
\midrule
{Sex} \\
\midrule
\hspace{0.5em} Female & 2731 & \textbf{81.9} (80.3--83.6) & \textbf{74.9} (72.5--77.6) & 81.8 (80.1--83.5) & 74.7 (72.0--77.3) \\
\hspace{0.5em} Male & 2711 & 81.3 (79.5--82.9) & 81.7 (79.6--83.6) & \textbf{82.9} (81.2--84.3) & \textbf{83.1} (81.0--85.1) \\
\midrule
{Race/ethnicity} \\
\midrule
\hspace{0.5em} Hispanic & 1649 & 82.7 (80.7--84.8) & 76.8 (73.6--79.9) & \textbf{83.3} (81.1--85.2) & \textbf{77.1} (74.0--80.4) \\
\hspace{0.5em} White & 1569 & 80.4 (78.3--82.5) & 76.1 (72.9--79.3) & \textbf{81.5} (79.3--83.4) & \textbf{77.7} (74.7--80.5)  \\
\hspace{0.5em} Black & 846 & 81.7 (78.8--84.5) & 81.8 (78.1--85.0) & \textbf{82.4} (79.5--84.9) & \textbf{82.6} (79.1--85.8) \\
\hspace{0.5em} Asian & 153 & 81.3 (74.3--87.9) & 81.8 (73.5--89.3) & \textbf{84.1} (77.9--89.9) & \textbf{83.8} (76.2--90.2) \\
\hspace{0.5em} Other & 457 & 81.5 (77.8--85.2) & 78.9 (74.0--83.8) & \textbf{82.4} (78.6--86.3) & \textbf{80.9} (76.1--85.4) \\
\hspace{0.5em} Unknown & 768 & 83.3 (80.4--86.1) & 83.4 (79.9--87.0) & \textbf{83.9} (80.9--86.6) & \textbf{83.5} (79.8--87.1) \\
\midrule
{Clinical context} \\
\midrule
\hspace{0.5em} Emergency & 1971 & 82.2 (80.3--84.1) & 77.6 (74.8--80.2) & \textbf{83.5} (81.8--85.3) & \textbf{78.8} (76.1--81.4) \\
\hspace{0.5em} Inpatient & 2203 & 79.7 (77.8--81.5) & 82.5 (80.4--84.5) & \textbf{80.0} (78.2--81.7) & \textbf{82.8} (80.7--84.7) \\
\hspace{0.5em} Outpatient & 1059 & 79.5 (76.2--82.6) & 64.5 (59.0--70.2) & \textbf{80.5} (77.2--83.7) & \textbf{67.1} (61.5--72.6) \\
\hspace{0.5em} Procedural & 209 & 79.5 (73.2--85.3) & 76.6 (67.1--85.3) & \textbf{82.2} (75.9--87.4) & \textbf{80.7} (72.4--88.2) \\
\bottomrule
\end{tabular}
}
\end{table}

\subsubsection{Estimated entry-wise functions}
Let
\[
\Big \{ \hat{\nu},\ \hat{\bm \gamma},\ ( \hat{\alpha}_{j,k})_{j=1, k=1}^{j = J, k=K-1}  \Big \}
\] be a solution to the optimization problem \eqref{eqn:def:opt}. To ensure identifiability, we reconstruct the estimated entry-wise function for each predictor $j$ as
\begin{equation}
\label{def:f_hat}
\hat{f}_j(x) := \sum_{k=1}^{K-1} \hat{\alpha}_{j,k} b_k(x) - \sum_{k=1}^{K-1} \int_0^1 \hat{\alpha}_{j,k} b_{j,k}(x) dx,\qquad j=1,\ldots,J.
\end{equation}
This centering constraint ensures that $\int_0^1 \hat{f}_j(x) dx = 0$, thereby preventing arbitrary constant shifts across components and enabling a consistent interpretation of nonlinear effects.

\begin{figure}[!htbp]
\centering
\subfloat[]{
\includegraphics[width=6.8cm]{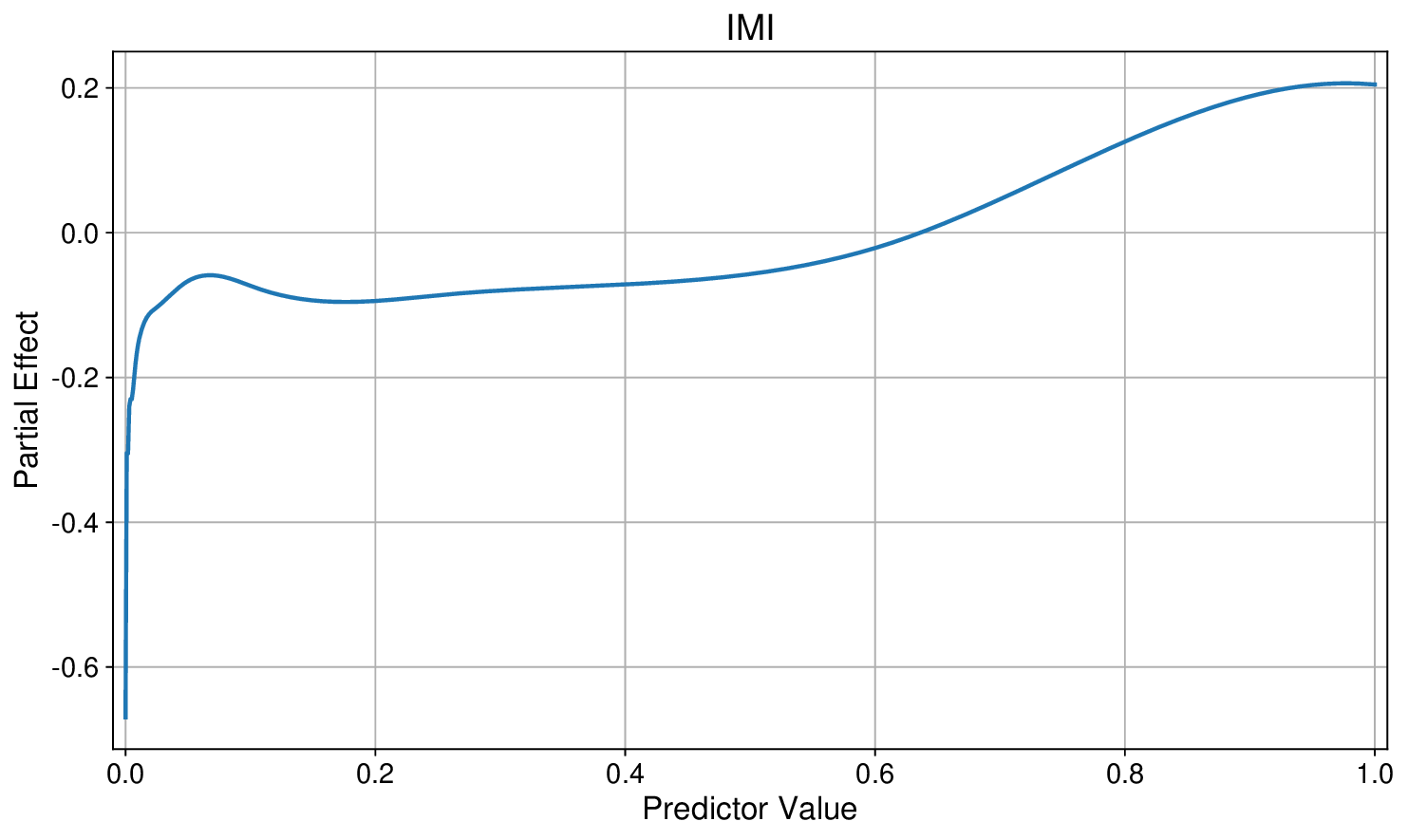}
\label{fig:subA}
}
\hfill
\subfloat[]{
\includegraphics[width=6.8cm]{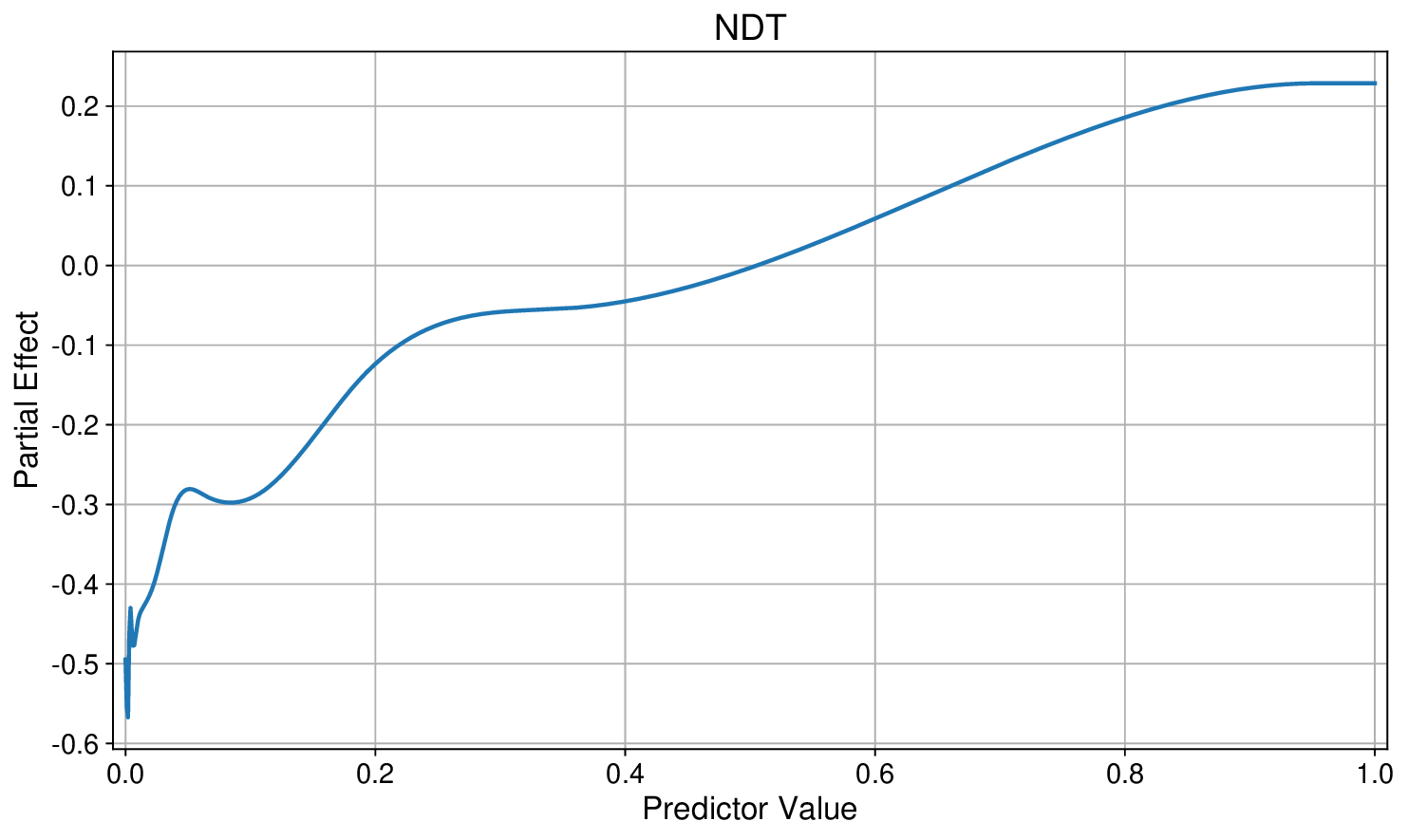}
\label{fig:subB}
}

\vspace{0.3cm}

\subfloat[]{
\includegraphics[width=6.8cm]{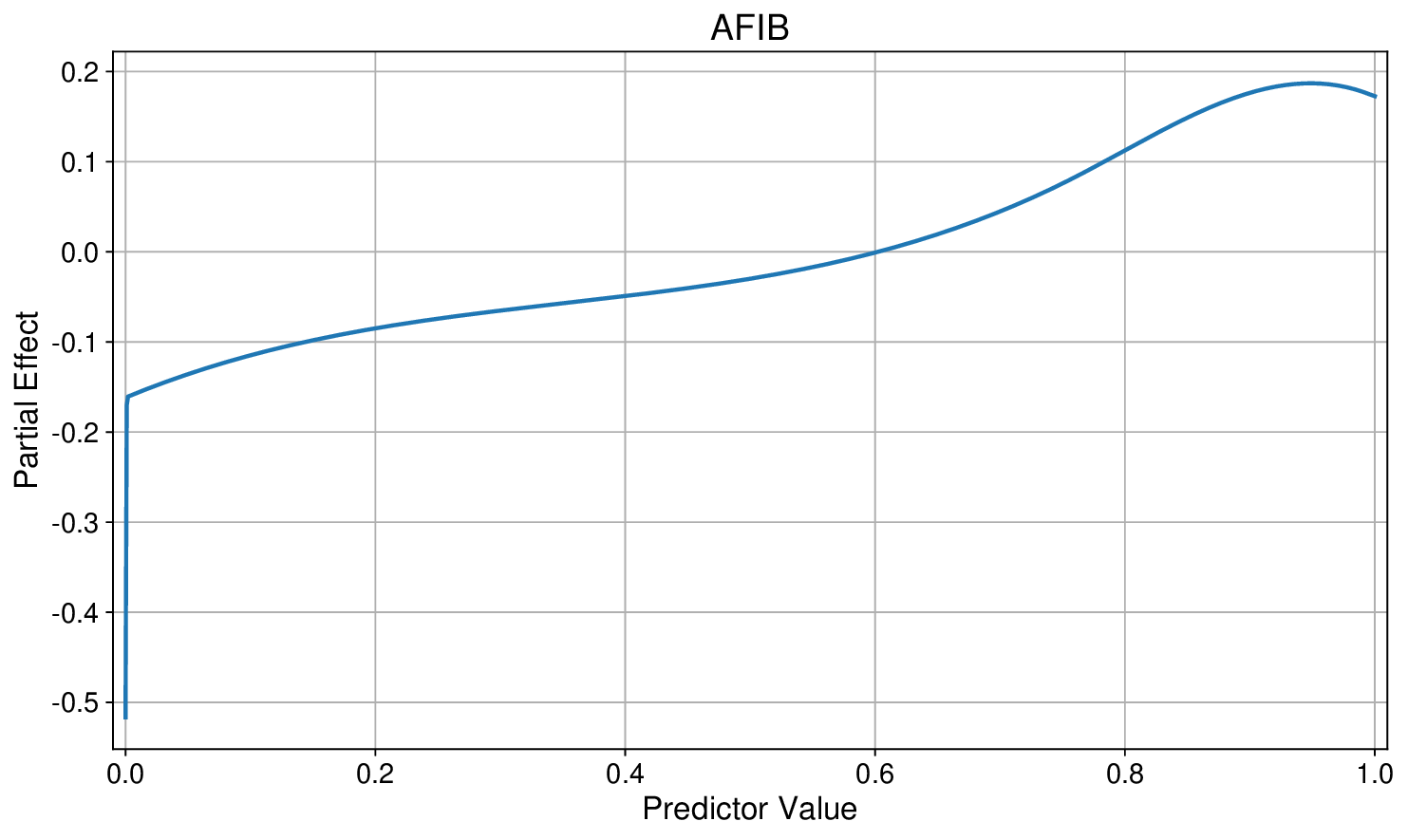}
\label{fig:subC}
}
\hfill
\subfloat[]{
\includegraphics[width=6.8cm]{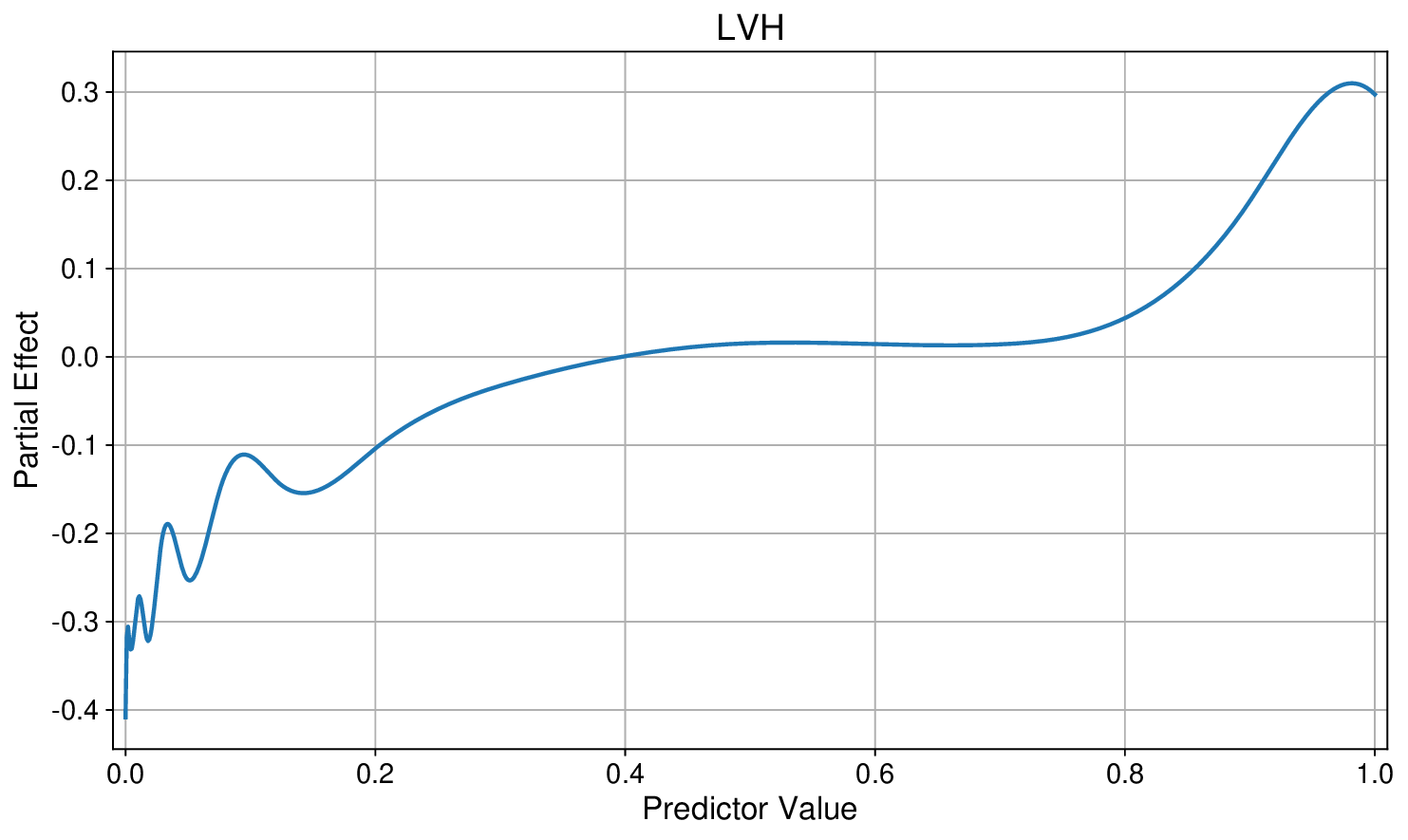}
\label{fig:subD}
}

\caption{Examples of estimated entry-wise functions in the proposed additive model.  ``Partial Effect'' refers to $\hat{f}_j(x)$ defined in \eqref{def:f_hat}, and ``Predictive Value'' refers to the corresponding $\sigma\{\hat{h}_j(\mathbf X)\}$, where $\hat{h}_j(\mathbf X)$ is defined in \eqref{eqn:def_hat_h}. The titles IMI, NDT, AFIB and LVH denote inferior myocardial infarction, non-diagnostic T abnormalities, atrial fibrillation, and left ventricular hypertrophy, respectively. 
}
\label{fig:entry_wise_esitimation}
\end{figure}

Recall that the predictors used in the additive model are derived from an ECG foundation model, where each predictor represents a model-estimated probability of a corresponding traditional ECG diagnoses. To illustrate the partial effects of these diagnoses, Figure \ref{fig:entry_wise_esitimation} presents four examples of estimated functions $\hat{f}_j(x)$ corresponding to inferior myocardial infarction (IMI), non-diagnostic T abnormalities (NDT), atrial fibrillation (AFIB), and left ventricular hypertrophy (LVH).

%Across these examples, the partial effect generally increases with the predictor value, reflecting the underlying risk associated with each traditional ECG abnormality. This monotonic trend aligns with clinical intuition: higher predicted probabilities of conventional ECG findings tend to indicate a higher likelihood of SHD. Notably, the observed nonlinear patterns suggest why subtle patterns of SHD may be difficult to detect visually from raw ECG signals. An important implication is that incorporating nonlinear decision rules into future ECG-based clinical guidance may help improve early detection of SHD.

Across these examples, the partial effect tends to increase with the predictor value, which is broadly consistent with the clinical intuition that higher predicted probabilities of traditional ECG abnormalities may be associated with a greater likelihood of SHD. The presence of nonlinear relationships may also help explain why some subtle ECG patterns related to SHD are challenging to recognize visually. While preliminary, these observations suggest that incorporating certain nonlinear decision components into future ECG-based clinical guidance may help support earlier detection of SHD.

\section{Discussion}
\label{sec:dis}
This study presents a simple yet effective framework that integrates an interpretable additive modeling strategy with predictors derived from an ECG foundation model for the detection of structural heart disease (SHD) from standard 12-lead ECGs. Using a benchmark dataset comprising more than 80,000 ECGs, the proposed approach demonstrated consistent improvements over the Columbia mini model \citep{poterucha2025detecting}, achieving gains of 0.98\%, 1.01\%, and 1.41\% in AUROC, AUPRC, and F1 score, respectively. In addition, experiments across varying training sample sizes indicate improved data efficiency: the proposed method achieved performance slightly better than the Columbia mini model using only 30\% of the training data. Subgroup analyses across heterogeneous patient subpopulations and care environments further demonstrated the robustness of the proposed approach. Importantly, each predictor within the additive model corresponds to a clinically meaningful latent predictor. The estimated entry-wise functions reveal nonlinear associations between risks of these predictors and SHD. Together, these properties position the method as an interpretable and practically applicable tool capable of supporting transparent decision-making in clinical workflows.

This work introduces a new paradigm for detecting SHD from ECG data. Traditionally, SHD has not been considered diagnosable via ECG because the associated ECG patterns might be subtle and imperceptible to human interpretation \citep{siontis2021artificial}. Recent advances demonstrated that such patterns can be uncovered by training deep learning models on large paired ECG–ECHO datasets \citep{attia2019screening, cohen2021electrocardiogram, poterucha2025detecting, diao2025speed}; however, these methods operate as end-to-end black-box systems, limiting transparency and clinical interpretability. The proposed framework represents a conceptual shift: instead of using deep learning solely for end-to-end prediction, the output of acceptable deep learning methods are transformed into interpretable predictors and modeled using an additive structure. This formulation preserves the predictive signal captured by modern artificial intelligence (AI) models while enabling explicit visualization and interpretation of each predictor’s contribution, offering a more explainable and clinically actionable approach to ECG-based SHD detection.

Building on these findings, this work also highlights the complementary roles of statistics and AI, as recently emphasized by \citet{redman2024ai}. While modern AI models have demonstrated strong predictive capabilities, they often sacrifice transparency and interpretability. Efforts in explainable AI aim to address this gap by designing models that not only make accurate predictions but also provide interpretable rationales \citep{broniatowski2021psychological}. However, such transparency may come with trade-offs in predictive accuracy or computational efficiency \citep{min2024applied}. Ongoing research seeks to address these drawbacks \citep{longo2024explainable, liu2025kan}. The proposed framework contributes to this direction by integrating clinically recognized predictors, which are already being adopted in practice, with a classical statistical modeling approach. The results in this paper demonstrate that it is possible to retain much of the predictive benefit of modern AI while preserving interpretability, suggesting a potential pathway for future applied statistical methodology in the AI era.

From a clinical perspective, the estimated smooth functions reveal nonlinear associations between risk of  traditional ECG diagnoses and SHD. Although SHD-related electrical signatures are often visually imperceptible, the foundation-model predictors reflect interpretable ECG constructs commonly used in clinical workflow. The observed nonlinear structure suggests that established ECG criteria may contain latent diagnostic information relevant to SHD that has not been systematically incorporated into clinical guidance. These findings motivate future work aimed at translating the estimated effects into clinically actionable decision thresholds, risk equations, or screening guidelines for ECG-based early detection of SHD.

Finally, this study has several limitations. First, SHD encompasses multiple phenotypes, yet the present analysis focuses on a composite outcome. Although this follows practice established in prior work \citep{ulloa2022rechommend, poterucha2025detecting}, it remains unknown whether the proposed framework generalizes consistently across individual SHD subtype labels. Second, the evaluation is conducted using a single ECG foundation model. More recent models trained on larger or more diverse datasets \citep{diao2025speed} may provide improved latent predictors and could further enhance performance. Third, the empirical analysis is based on data from a single center without external or multi-center validation, which limits the assessment of model generalizability. These limitations represent active directions for future research.

\section{Conclusion}
\label{sec:con}
%In summary, this study proposes a framework that integrates foundation-model–derived ECG representations with an interpretable additive modeling structure to detect structural heart disease from standard 12-lead ECGs. The approach achieves strong predictive performance while retaining transparency in how individual predictors contribute to risk estimation, demonstrating that interpretability and modern AI need not be mutually exclusive. By bridging statistical modeling principles with advances in ECG foundation models, the framework offers a practical and clinically aligned direction for AI-assisted cardiac screening. Future work will evaluate performance across individual SHD subtypes, additional foundation model architectures, and external multi-center cohorts to further assess generalizability and clinical utility.

In summary, this study proposes a novel framework that integrates interpretable foundation-model–derived ECG predictors with an additive modeling structure to detect SHD from standard 12-lead ECGs. Experiments show that the approach maintains predictive accuracy and robustness while providing transparency in how individual predictors contribute to risk estimation, illustrating an approach that is both interpretable and practically applicable. These findings also present an example that statistics and modern AI are complementary. Future work will evaluate performance across individual SHD subtypes, additional foundation model architectures, and external multi-center datasets to further assess generalizability and clinical utility.

\bibliographystyle{rss} 
\bibliography{reference}

\begin{thebibliography}{40}
\expandafter\ifx\csname natexlab\endcsname\relax\def\natexlab#1{#1}\fi
\expandafter\ifx\csname url\endcsname\relax
  \def\url#1{\texttt{#1}}\fi
\expandafter\ifx\csname urlprefix\endcsname\relax\def\urlprefix{URL: }\fi

\bibitem[{Attia et~al.(2019)}]{attia2019screening}
Attia, Z.~I. et~al. (2019) Screening for cardiac contractile dysfunction using
  an artificial intelligence--enabled electrocardiogram.
\newblock \textit{Nature Med.}, \textbf{25}, 70--74.

\bibitem[{Behrouz et~al.(2024)Behrouz, Santacatterina and
  Zabih}]{behrouz2024chimera}
Behrouz, A., Santacatterina, M. and Zabih, R. (2024) Chimera: Effectively
  modeling multivariate time series with 2-dimensional state space models.
\newblock In \textit{The Thirty-eighth Annual Conference on Neural Information
  Processing Systems}.
\newblock \urlprefix\url{https://openreview.net/forum?id=ncYGjx2vnE}.

\bibitem[{Broniatowski and Broniatowski(2021)}]{broniatowski2021psychological}
Broniatowski, D.~A. and Broniatowski, D.~A. (2021) \textit{Psychological
  foundations of explainability and interpretability in artificial
  intelligence}, vol.~4.
\newblock US Department of Commerce, National Institute of Standards and
  Technology.

\bibitem[{Cheng et~al.(2023)Cheng, Zhuang, Zhuang, Xie and Guo}]{cheng2023msw}
Cheng, R., Zhuang, Z., Zhuang, S., Xie, L. and Guo, J. (2023) Msw-transformer:
  Multi-scale shifted windows transformer networks for 12-lead ecg
  classification.
\newblock \textit{arXiv preprint arXiv:2306.12098}.

\bibitem[{Cohen-Shelly et~al.(2021)Cohen-Shelly, Attia, Friedman, Ito,
  Essayagh, Ko, Murphree, Michelena, Enriquez-Sarano, Carter
  et~al.}]{cohen2021electrocardiogram}
Cohen-Shelly, M., Attia, Z.~I., Friedman, P.~A., Ito, S., Essayagh, B.~A., Ko,
  W.-Y., Murphree, D.~H., Michelena, H.~I., Enriquez-Sarano, M., Carter, R.~E.
  et~al. (2021) Electrocardiogram screening for aortic valve stenosis using
  artificial intelligence.
\newblock \textit{European Heart Journal}, \textbf{42}, 2885--2896.

\bibitem[{Coppola et~al.(2024)Coppola, Savardi, Massussi, Adamo, Metra and
  Signoroni}]{coppola2024hubert}
Coppola, E., Savardi, M., Massussi, M., Adamo, M., Metra, M. and Signoroni, A.
  (2024) Hubert-ecg: a self-supervised foundation model for broad and scalable
  cardiac applications.
\newblock \textit{medRxiv}, 2024--11.

\bibitem[{d'Arcy et~al.(2016)d'Arcy, Coffey, Loudon, Kennedy, Pearson-Stuttard,
  Birks, Frangou, Farmer, Mant, Wilson et~al.}]{d2016large}
d'Arcy, J.~L., Coffey, S., Loudon, M.~A., Kennedy, A., Pearson-Stuttard, J.,
  Birks, J., Frangou, E., Farmer, A.~J., Mant, D., Wilson, J. et~al. (2016)
  Large-scale community echocardiographic screening reveals a major burden of
  undiagnosed valvular heart disease in older people: the oxvalve population
  cohort study.
\newblock \textit{European heart journal}, \textbf{37}, 3515--3522.

\bibitem[{Diao et~al.(2025)Diao, Xu, Cheng, Zhou, Liu, Huo, Lu, Huang, He, Liu
  et~al.}]{diao2025speed}
Diao, X., Xu, W., Cheng, H., Zhou, Y., Liu, Y., Huo, Y., Lu, J., Huang, J., He,
  J., Liu, F. et~al. (2025) Speed-tr: a self-distilled and pre-trained
  transformer model for enhanced ecg detection of tricuspid regurgitation.
\newblock \textit{npj Digital Medicine}, \textbf{8}, 650.

\bibitem[{Elias and Finer(2025)}]{elias2025echonext}
Elias, P. and Finer, J. (2025) Echonext: A dataset for detecting
  echocardiogram-confirmed structural heart disease from ecgs.

\bibitem[{Fan et~al.(2014)Fan, Ma and Dai}]{fan2014nonparametric}
Fan, J., Ma, Y. and Dai, W. (2014) Nonparametric independence screening in
  sparse ultra-high-dimensional varying coefficient models.
\newblock \textit{Journal of the American Statistical Association},
  \textbf{109}, 1270--1284.

\bibitem[{Hannun et~al.(2019)}]{hannun2019cardiologist}
Hannun, A.~Y. et~al. (2019) Cardiologist-level arrhythmia detection and
  classification in ambulatory electrocardiograms using a deep neural network.
\newblock \textit{Nature Med.}, \textbf{25}, 65--69.

\bibitem[{Hastie and Tibshirani(1990)}]{hastie1990generalized}
Hastie, T. and Tibshirani, R. (1990) \textit{Generalized Additive Models},
  vol.~43.
\newblock CRC Press.

\bibitem[{Hata et~al.(2020)Hata, Seo, Nakayama, Iwasaki, Ohkawauchi and
  Ohya}]{hata2020classification}
Hata, E., Seo, C., Nakayama, M., Iwasaki, K., Ohkawauchi, T. and Ohya, J.
  (2020) Classification of aortic stenosis using ecg by deep learning and its
  analysis using grad-cam.
\newblock In \textit{2020 42nd Annual International Conference of the IEEE
  Engineering in Medicine \& Biology Society (EMBC)}, 1548--1551. IEEE.

\bibitem[{He et~al.(2022)He, Chen, Xie, Li, Doll{\'a}r and
  Girshick}]{he2021masked}
He, K., Chen, X., Xie, S., Li, Y., Doll{\'a}r, P. and Girshick, R. (2022)
  Masked autoencoders are scalable vision learners.
\newblock In \textit{Proc. IEEE/CVF Conf. Comput. Vis. Pattern Recognit.,
  \textnormal{2022, pp. 16000--16009}}.

\bibitem[{Huang et~al.(2010)Huang, Horowitz and Wei}]{huang2010variable}
Huang, J., Horowitz, J.~L. and Wei, F. (2010) Variable selection in
  nonparametric additive models.
\newblock \textit{Annals of Statistics}, \textbf{38}, 2282.

\bibitem[{Jiang et~al.(2024)Jiang, Huang, Cao, Xu, Zeng, Chen, Zhang and
  Wang}]{jiang2024self}
Jiang, A., Huang, C., Cao, Q., Xu, Y., Zeng, Z., Chen, K., Zhang, Y. and Wang,
  Y. (2024) Self-supervised anomaly detection pretraining enhances long-tail
  ecg diagnosis.
\newblock \textit{arXiv preprint arXiv:2408.17154}.

\bibitem[{Kwon et~al.(2020)Kwon, Lee, Jeon, Lee, Kim, Park, Oh and
  Lee}]{kwon2020deep}
Kwon, J.-M., Lee, S.~Y., Jeon, K.-H., Lee, Y., Kim, K.-H., Park, J., Oh, B.-H.
  and Lee, M.-M. (2020) Deep learning--based algorithm for detecting aortic
  stenosis using electrocardiography.
\newblock \textit{Journal of the American Heart Association}, \textbf{9},
  e014717.

\bibitem[{Li et~al.(2025)Li, Aguirre, Junior, Jin, Liu, Zhong, Sun, Clifford,
  Brandon~Westover and Hong}]{li2025electrocardiogram}
Li, J., Aguirre, A.~D., Junior, V.~M., Jin, J., Liu, C., Zhong, L., Sun, C.,
  Clifford, G., Brandon~Westover, M. and Hong, S. (2025) An electrocardiogram
  foundation model built on over 10 million recordings.
\newblock \textit{NEJM AI}, \textbf{2}, AIoa2401033.

\bibitem[{Liu et~al.(2025)Liu, Wang, Vaidya, Ruehle, Halverson, Soljacic, Hou
  and Tegmark}]{liu2025kan}
Liu, Z., Wang, Y., Vaidya, S., Ruehle, F., Halverson, J., Soljacic, M., Hou,
  T.~Y. and Tegmark, M. (2025) {KAN}: Kolmogorov{\textendash}arnold networks.
\newblock In \textit{The Thirteenth International Conference on Learning
  Representations}.
\newblock \urlprefix\url{https://openreview.net/forum?id=Ozo7qJ5vZi}.

\bibitem[{Longo et~al.(2024)Longo, Brcic, Cabitza, Choi, Confalonieri, Del~Ser,
  Guidotti, Hayashi, Herrera, Holzinger et~al.}]{longo2024explainable}
Longo, L., Brcic, M., Cabitza, F., Choi, J., Confalonieri, R., Del~Ser, J.,
  Guidotti, R., Hayashi, Y., Herrera, F., Holzinger, A. et~al. (2024)
  Explainable artificial intelligence (xai) 2.0: A manifesto of open challenges
  and interdisciplinary research directions.
\newblock \textit{Information Fusion}, \textbf{106}, 102301.

\bibitem[{McKeen et~al.(2024)McKeen, Masood, Toma, Rubin and
  Wang}]{mckeen2024ecg}
McKeen, K., Masood, S., Toma, A., Rubin, B. and Wang, B. (2024) Ecg-fm: An open
  electrocardiogram foundation model.
\newblock \textit{arXiv preprint arXiv:2408.05178}.

\bibitem[{Mensah et~al.(2023)Mensah, Fuster, Murray, Roth,
  of~Cardiovascular~Diseases and Collaborators}]{mensah2023global}
Mensah, G.~A., Fuster, V., Murray, C.~J., Roth, G.~A.,
  of~Cardiovascular~Diseases, G.~B. and Collaborators, R. (2023) Global burden
  of cardiovascular diseases and risks, 1990-2022.
\newblock \textit{Journal of the American College of Cardiology}, \textbf{82},
  2350--2473.

\bibitem[{Min et~al.(2024)Min, Song, Zheng, King, Deng and
  Hong}]{min2024applied}
Min, J., Song, X., Zheng, S., King, C.~B., Deng, X. and Hong, Y. (2024) Applied
  statistics in the era of artificial intelligence: A review and vision.
\newblock \textit{arXiv preprint arXiv:2412.10331}.

\bibitem[{Na et~al.(2024)Na, Park, Tae and Joo}]{na2024guiding}
Na, Y., Park, M., Tae, Y. and Joo, S. (2024) Guiding masked representation
  learning to capture spatio-temporal relationship of electrocardiogram.
\newblock In \textit{The Twelfth International Conference on Learning
  Representations}.
\newblock \urlprefix\url{https://openreview.net/forum?id=WcOohbsF4H}.

\bibitem[{Poterucha et~al.(2025)Poterucha, Jing, Ricart, Adjei-Mosi, Finer,
  Hartzel, Kelsey, Long, Rocha, Ruhl et~al.}]{poterucha2025detecting}
Poterucha, T.~J., Jing, L., Ricart, R.~P., Adjei-Mosi, M., Finer, J., Hartzel,
  D., Kelsey, C., Long, A., Rocha, D., Ruhl, J.~A. et~al. (2025) Detecting
  structural heart disease from electrocardiograms using ai.
\newblock \textit{Nature}, \textbf{644}, 221--230.

\bibitem[{Redman and Hoerl(2024)}]{redman2024ai}
Redman, T.~C. and Hoerl, R.~W. (2024) Ai and statistics: Perfect together.
\newblock \textit{MIT Sloan Management Review (Online)}, 1--4.

\bibitem[{Ribeiro et~al.(2020)}]{ribeiro2020automatic}
Ribeiro, A.~H. et~al. (2020) Automatic diagnosis of the 12-lead ecg using a
  deep neural network.
\newblock \textit{Nature Commun.}, \textbf{11}, 1760.

\bibitem[{Shi et~al.(2023)Shi, Wang and Fox}]{shi2023sequence}
Shi, J., Wang, K.~A. and Fox, E. (2023) Sequence modeling with multiresolution
  convolutional memory.
\newblock In \textit{International Conference on Machine Learning},
  31312--31327. PMLR.

\bibitem[{Siontis et~al.(2021)Siontis, Noseworthy, Attia and
  Friedman}]{siontis2021artificial}
Siontis, K.~C., Noseworthy, P.~A., Attia, Z.~I. and Friedman, P.~A. (2021)
  Artificial intelligence-enhanced electrocardiography in cardiovascular
  disease management.
\newblock \textit{Nature Rev. Cardiol.}, \textbf{18}, 465--478.

\bibitem[{Somani et~al.(2021)}]{somani2021deep}
Somani, S. et~al. (2021) Deep learning and the electrocardiogram: review of the
  current state-of-the-art.
\newblock \textit{EP Europace}, \textbf{23}, 1179--1191.

\bibitem[{Stone(1985)}]{stone1985additive}
Stone, C.~J. (1985) Additive regression and other nonparametric models.
\newblock \textit{The annals of Statistics}, 689--705.

\bibitem[{Strodthoff(2024)}]{strodthoffopen}
Strodthoff, N. (2024) Open science to foster progress in automatic ecg
  analysis: Status and future directions.

\bibitem[{Strodthoff et~al.(2021)Strodthoff, Wagner, Schaeffter and
  Samek}]{strodthoff2020deep}
Strodthoff, N., Wagner, P., Schaeffter, T. and Samek, W. (2021) Deep learning
  for ecg analysis: Benchmarks and insights from ptb-xl.
\newblock \textit{IEEE J. Biomed. Health Inform.}, \textbf{25}, 1519--1528.

\bibitem[{Tsao et~al.(2023)Tsao, Aday, Almarzooq, Anderson, Arora, Avery,
  Baker-Smith, Beaton, Boehme, Buxton et~al.}]{tsao2023heart}
Tsao, C.~W., Aday, A.~W., Almarzooq, Z.~I., Anderson, C.~A., Arora, P., Avery,
  C.~L., Baker-Smith, C.~M., Beaton, A.~Z., Boehme, A.~K., Buxton, A.~E. et~al.
  (2023) Heart disease and stroke statistics—2023 update: a report from the
  american heart association.
\newblock \textit{Circulation}, \textbf{147}, e93--e621.

\bibitem[{Ulloa-Cerna et~al.(2022)Ulloa-Cerna, Jing, Pfeifer, Raghunath, Ruhl,
  Rocha, Leader, Zimmerman, Lee, Steinhubl et~al.}]{ulloa2022rechommend}
Ulloa-Cerna, A.~E., Jing, L., Pfeifer, J.~M., Raghunath, S., Ruhl, J.~A.,
  Rocha, D.~B., Leader, J.~B., Zimmerman, N., Lee, G., Steinhubl, S.~R. et~al.
  (2022) rechommend: an ecg-based machine learning approach for identifying
  patients at increased risk of undiagnosed structural heart disease detectable
  by echocardiography.
\newblock \textit{Circulation}, \textbf{146}, 36--47.

\bibitem[{Wagner et~al.(2020)}]{wagner2020ptb}
Wagner, P. et~al. (2020) Ptb-xl, a large publicly available electrocardiography
  dataset.
\newblock \textit{Sci. Data}, \textbf{7}, 1--15.

\bibitem[{Wang et~al.(2023)Wang, Ma, Liu, Fan, Hu, of~the Report~on
  Cardiovascular~Health and in~China}]{wang2023summary}
Wang, Z., Ma, L., Liu, M., Fan, J., Hu, S., of~the Report~on
  Cardiovascular~Health, W.~C. and in~China, D. (2023) Summary of the 2022
  report on cardiovascular health and diseases in china.
\newblock \textit{Chinese Medical Journal}, \textbf{136}, 2899--2908.

\bibitem[{Zhang et~al.(2023)Zhang, Saab, Poli, Dao, Goel and
  Re}]{zhang2023effectively}
Zhang, M., Saab, K.~K., Poli, M., Dao, T., Goel, K. and Re, C. (2023)
  Effectively modeling time series with simple discrete state spaces.
\newblock In \textit{Proc. Int. Conf. Learn. Representations}.

\bibitem[{Zhou et~al.(2025{\natexlab{a}})Zhou, Diao, Huo, Liu, Sun, Fan and
  Zhao}]{zhou2025enhancing}
Zhou, Y., Diao, X., Huo, Y., Liu, Y., Sun, Z., Fan, X. and Zhao, W.
  (2025{\natexlab{a}}) Enhancing automatic multilabel diagnosis of
  electrocardiogram signals: A masked transformer approach.
\newblock \textit{Computers in Biology and Medicine}, \textbf{196}, 110674.

\bibitem[{Zhou et~al.(2025{\natexlab{b}})Zhou, Yang, Fan and
  Zhao}]{zhou2025bridging}
Zhou, Y., Yang, Y., Fan, X. and Zhao, W. (2025{\natexlab{b}}) Bridging
  performance gaps for foundation models: A post-training strategy for
  ecgfounder.
\newblock \textit{arXiv preprint arXiv:2509.12991}.

\end{thebibliography}

\end{document}